\makeatletter \renewcommand\@biblabel[1]{{#1}. } \makeatother
\documentstyle[epsfig,12pt,doublespace,overcite]{article} 
\newcommand{\ah}{$\alpha_{h}$}
\newcommand{\as}{$\alpha_{s}$}

\newcommand{\metro}{p = \left\{ \begin{array}{ll}
				1 & \delta H \leq 0 \\
				\exp(-\delta H/T) & \mathrm{otherwise} \\
				\end{array}
			\right. }

\newcommand{\hamiltonian}{\delta H = \delta U + \left\{ \begin{array}{ll}
				C \left(R_{min}-R\right) & R < R_{min} \\
				0  & R_{min} \leq R \leq R_{max} \\
				C \left(R-R_{max}\right) & R > R_{max} \\				
				\end{array}
			\right. }

\begin{document}

\title{\textbf{Excluded Volume in Protein Sidechain Packing}}

\author{Edo Kussell, Jun Shimada, Eugene I. Shakhnovich$^{\star}$}

\date{\today}
 
\maketitle
\begin{center}
\vspace{0.25in}
\large
\textbf{Running title:} Sidechain Packing in Proteins\\
\normalsize
\vspace{0.55in}
36 pages (including figure/table captions), 7 figures, 2 tables.\\  
\vspace{0.5in}
\begin{spacing}{1}
Department of Chemistry and Chemical Biology \\
Harvard University \\
12 Oxford Street  \\
Cambridge MA 02138\\
\vspace{0.5in}
$^\star$corresponding author \\
tel: 617-495-4130 \\
fax: 617-496-5948 \\
email: eugene@belok.harvard.edu \\
\vspace{0.5in}

\end{spacing}
\end{center}

\newpage
\noindent
\textbf{Abstract}

\noindent
The excluded volume occupied by protein sidechains and the requirement
of high packing density in the protein interior should severely limit
the number of sidechain conformations compatible with a given native
backbone.  To examine the relationship between sidechain geometry and
sidechain packing, we use an all-atom Monte Carlo simulation to sample
the large space of sidechain conformations.  We study three models of
excluded volume and use umbrella sampling to effectively explore the
entire space.  We find that while excluded volume constraints reduce
the size of conformational space by many orders of magnitude, the
number of allowed conformations is still large.  An average repacked
conformation has 20\% of its $\chi$ angles in a non-native state, a
marked reduction from the expected 67\% in the absence of excluded
volume.  Interestingly, well-packed conformations with up to 50\%
non-native $\chi$'s exist.  The repacked conformations have native
packing density as measured by a standard Voronoi procedure.  Entropy
is distributed non-uniformly over positions, and we partially explain
the observed distribution using rotamer probabilities derived from the
pdb database~\cite{kn:dunbrack_cohen}.  In several cases, native
rotamers which occur infrequently in the pdb database are seen with
high probability in our simulation, indicating that sequence-specific
excluded volume interactions can stabilize rotamers that are rare for
a given backbone.  In spite of our finding that 65\% of the native
rotamers and 85\% of $\chi_{1}$ angles can be correctly predicted on
the basis of excluded volume only, 95\% of positions can accomodate
more than 1 rotamer in simulation.  We estimate that in order to quench
the sidechain entropy observed in the presence of excluded volume
interactions, other interactions (hydrophobic, polar, electrostatic)
must provide an additional stabilization of at least 0.6 kT per
residue in order to single out the native state.

\vspace{1in}

\newpage
\section*{Introduction}
One remarkable characteristic of proteins is that sidechains
comprising the hydrophobic core are as closely packed as organic
crystals~\cite{kn:fersht}.  In most proteins with known 3D structures,
the core residues are rarely disordered and adopt one of a small
number of alternative conformations.  This almost unique packing is
effected by a combination of steric interactions (excluded volume
effects) and energetic stabilization (hydrophobic, polar, and charge
interactions).  The proportions by which these interactions contribute
to the overall stability is unknown, but several studies suggest that
steric and hydrophobic interactions are of primary
importance~\cite{kn:lee_subbiah,kn:ponder_richards}.

Sidechain packing has been studied intensively for various reasons,
notably: 1) It is thought to be a crucial piece of the protein folding
puzzle~\cite{kn:richards_lim,kn:shakh_finkel}; 2) The selection of
protein sequences through evolution may have been influenced by how
well a sequence can be packed for a particular
fold~\cite{kn:ponder_richards}; 3) Accurate packing algorithms are
necessary for complete protein structure prediction; and 4) Existing
threading and homology modeling algorithms may be significantly
improved by a better understanding of how sidechains are stabilized in
the core~\cite{kn:bower_dunbrack,kn:mirny_threading}.  A plethora of
computational methods for modeling protein sidechains and energetics
exist~\cite{kn:lee_subbiah,kn:bower_dunbrack,kn:petrella_karplus,kn:mendes_soares},
and yet which ones best capture the underlying physics is unclear.
This is due, in part, to the use of energy functions containing many
different types of interactions.

In this paper, we examine the role of excluded volume in isolation of
all other types of interactions.  This allows us to address the
importance of geometry in determining the native state of protein
sidechains.  We use an all-atom, rotamer-based model of a protein to
obtain repacked conformations of its interior sidechains.  By choosing
the most realistic representation of sidechain and backbone
geometries, we eliminate errors encountered in coarse-grained models.

Because there is no consensus on how excluded volume interactions
should be modeled in the context of proteins, we consider three
different models.  The simplest model treats all heavy atoms as hard
spheres which are not allowed to overlap.  The second model further
restricts the sterically allowed space by adding a second shell around
the hard spheres which tolerates only a limited number of overlaps
throughout the entire protein.  The third model uses a continuous
$r^{-12}$ potential, which is the repulsive contribution from a
Lennard-Jones potential.  Since a potential which decays faster than
our chosen $r^{-12}$ potential would allow more sidechain
conformations, and since a slower decaying function is generally not
used to model sterics~\cite{kn:mcquarrie}, the three we have chosen
should cover most of the possibilities as far as models featuring
spherically symmetric, atom-based potentials.

For each model, we obtain the distribution of rms values for sidechain
conformations satisfying excluded volume constraints.  This amounts to
a full characterization of the ground-state ensemble of protein
sidechains subject to excluded volume effects only.  We find a vast
number of significantly different conformations with native-like
packing density that meet excluded volume constraints, implying that
excluded volume alone cannot stabilize the native conformation of
sidechains even when the backbone is fixed.  We provide an estimate
for the amount of additional stabilization that must be provided
through interactions other than sterics.  The results obtained using
three different models are consistent with each other, suggesting that
our main conclusions are model-independent.  We also examine the
correlation between our simulations and the distribution of rotamers
in the pdb database, and discuss implications for sidechain
prediction.  Our main goal is to elucidate the role of excluded volume
in establishing the ensemble of conformations compatible with the
folded state of a protein.

\section*{Methods}
\textbf{All-Atom Protein Representation.}  Three proteins were used in
this study: photoactive yellow protein (1F9I), subtilisin (1GCI), and
concanavalin (1NLS).  We chose these proteins because they have very
high resolution crystal structures, and are of different
secondary-structure classes: subtilisin is a 0.78 \AA\ resolution
structure, and is an all $\alpha$-helical protein; concanavalin (0.94
\AA\ resolution) is all $\beta$-sheet; and photoactive yellow protein
(1.1 \AA\ resolution) is $\alpha$/$\beta$.  All heavy atoms of the
proteins present in the crystal structure were represented.  Each
sidechain torsion was allowed to take on one of three rotamer
values~\cite{kn:dunbrack_cohen} plus a random noise of $\pm
10^{\circ}$.  Rotamer probabilities~\cite{kn:dunbrack_cohen} were not
used in the simulation.  At all times, the atoms of the entire
backbone were kept fixed in their original crystal structure
positions.

\noindent
\textbf{Models of sterics.} The excluded volume interactions of
protein sidechains were modeled in three ways:

1) In the \emph{hard-sphere model} atoms were treated as impenetrable
spheres of given radii.  These hard-sphere radii are necessarily
smaller than the van der Waals (vdW) radii, and were defined by
scaling the vdW radii by some factor \ah.  All atom-atom interactions
at distances smaller than the sum of the hard-sphere radii (``hard
clashes'') were strictly forbidden.  A pair of atoms $i$ and $j$
separated by a distance $r_{ij}$ is said to be a hard clash if $r_{ij}
< \alpha_{h} \left(r_{0}(i) + r_{0}(j)\right)$, where $r_{0}(i)$ is
the vdW radius of atom $i$.  The set of vdW radii determined
in~\cite{kn:gerstein_chothia_99} was used.  \ah\ was taken to be the
largest value such that all hard clashes in the native protein
conformation can be eliminated within the $\pm 10^{\circ}$ allowed at
each torsion; we found that \ah $= 0.76$ for concanavalin and
photoactive yellow protein, and $0.77$ for subtilisin.  The steric
energy, $U_h$, of a conformation in this model is given by the number
of hard clashes.  Only conformations with $U_h = 0$ are sterically
allowed in this model.

2) Since results could depend on the choice of \ah, we explored a
\emph{soft-sphere model} in which atoms consisted of two radii - the
hard-sphere radii fixed by \ah, and somewhat larger, soft-sphere
radii.  Soft radii were defined by scaling vdW radii by a parameter
\as, where $\alpha_{h} < \alpha_{s} \leq 1$.  A soft clash occurs when
$\alpha_{h} \left(r_{0}(i) + r_{0}(j)\right) < r_{ij} < \alpha_{s}
\left(r_{0}(i) + r_{0}(j)\right)$.  All hard clashes were forbidden as
before, while only a limited number of soft clashes are allowed over
the entire protein.  The steric energy, $U$, of a conformation is
given by $U = U_h + U_s$, where $U_h$ is the number of hard atom-atom
clashes, and $U_s$ is the number of soft atom-atom clashes above
threshold.  The threshold number of soft clashes allowed is a function
of \as, as described in Fig.~\ref{fig:alpha_range}.  Only
conformations with $U = 0$ are sterically allowed in this model.
Since van der Waals radii are strictly greater than excluded volume
radii, as demonstrated in numerous studies of gas phase organic
molecules~\cite{kn:mcquarrie}, \as\ must be strictly less than 1.0,
and we include \as $= 1.0$ in our plots only as an upper limit on the
radii.

3) The third model used the repulsive term of a \emph{Lennard-Jones}
potential.  For each pair of atoms we fit the LJ potential, $A/r^{12}
- 1/r^{6}$, so that the minimum coincides with the sum of the vdW
radii taken from~\cite{kn:gerstein_chothia_99}.  With this model, the
steric energy of a conformation is given by $U_{LJ} = \sum_{ij}
A_{ij}/r^{12}_{ij}$ where the sum is taken over all pairs of atoms,
and $A_{ij} = (r_{0}(i) + r_{0}(j))^{6}/2$. We considered a
conformation to be sterically allowed if its energy was less than or
equal to the energy of the native crystal structure.

\textbf{Monte Carlo Packing Algorithm.} Initial random sidechain
conformations were packed to a given rms interval $[R_{min},R_{max}]$
by a Metropolis Monte Carlo
simulation~\cite{kn:metropolis,kn:binder_92}.  At each step of the
simulation, a random position was selected and changed to another
rotamer.  The move was accepted with probability $p$ given by
\[\metro\] with \[\hamiltonian\] where $U$ is the steric
energy, $R$ is the rms, and $C = 10$.  All reported rms values were
computed over the subset of sidechains having more than 40\% occluded
surface area as defined in reference \citen{kn:occ_surface}, and at
least 1 rotatable $\chi$ angle.  All heavy atoms of a sidechain,
including the $C_{\beta}$ atom, entered into the rms calculation as in
reference \citen{kn:lee_subbiah}.  Prolines were kept fixed in their
crystal structure positions throughout this work.  The 40\% occluded
positions for each protein are given below.  Additionally, by visual
inspection we found positions whose $C_{\alpha}$-$C_{\beta}$ bonds
were pointing toward the core, and/or whose sidechains were surrounded
by other atoms on all sides.  These positions are indicated in bold
type.  \\ \\ Subtilisin: \\ (1)
\textbf{8I},\textbf{11V},17H,21L,22T,\textbf{26V},27K,\textbf{28V},\textbf{30V}
\\ (10)
\textbf{31L},\textbf{32D},\textbf{33T},\textbf{35I},36S,38H,40D,\textbf{41L},49F,50V
\\ (20)
\textbf{58D},62H,\textbf{64T},\textbf{65H},\textbf{66V},\textbf{69T},\textbf{70I},73L,79V,80L
\\ (30)
\textbf{82V},87E,\textbf{88L},89Y,\textbf{91V},\textbf{92K},\textbf{93V},94L,\textbf{104S},105I
\\ (40)
\textbf{109L},111W,117M,\textbf{119V},\textbf{121N},\textbf{122L},123S,130S,133L,137V
\\ (50)
139S,141T,145V,146L,\textbf{147V},\textbf{148V},\textbf{151S},\textbf{159I},160S,161Y
\\ (60)
165Y,\textbf{169M},\textbf{171V},\textbf{174T},175D,183F,\textbf{184S},185Q,186Y,\textbf{190L}
\\ (70)
191D,\textbf{192I},\textbf{193V},197V,199V,\textbf{201S},\textbf{202T},203Y,\textbf{214T},215S
\\ (80)
216M,\textbf{218T},\textbf{220H},\textbf{221V},\textbf{227L},\textbf{228V},229K,232N,237N,240I
\\ (90)
241R,243H,\textbf{244L},245K,256L,257Y,261L,\textbf{262V},263N,265E \\
(100)268T \\ \\ Concanavalin: \\ (1)
4I,5V,\textbf{7V},\textbf{8E},\textbf{9L},\textbf{10D},11T,14N,17I \\
(10)
\textbf{19D},\textbf{24H},\textbf{25I},\textbf{27I},\textbf{28D},29I,31S,\textbf{32V},\textbf{40W},\textbf{52I}
\\ (20)
\textbf{54Y},55N,\textbf{61L},62S,\textbf{65V},67Y,75V,79V,80D,\textbf{81L}
\\ (30)
\textbf{85L},\textbf{89V},90R,\textbf{91V},\textbf{93L},\textbf{94S},\textbf{96S},\textbf{97T},\textbf{102E},103T
\\ (40)
\textbf{104N},\textbf{105T},\textbf{106I},108S,\textbf{109W},\textbf{110S},\textbf{111F},113S,115L,117S
\\ (50)
\textbf{128F},130F,\textbf{133F},\textbf{134S},\textbf{140L},143Q,145D,148T,153N,154L
\\ (60)
\textbf{156L},\textbf{157T},164S,169S,170V,\textbf{172R},174L,175F,179V,\textbf{180H}
\\ (70)
\textbf{191F},195F,196T,\textbf{197F},\textbf{199I},\textbf{201S},208D,\textbf{210I},211A,\textbf{213F}
\\ (80)
\textbf{214I},215S,216N,\textbf{219S},\textbf{225S},229L,\textbf{230L},\textbf{232L},233F
\\ \\ Photoactive Yellow Protein: \\ (1)
6F,\textbf{11I},12E,\textbf{15L},18M,22Q,23L,\textbf{26L},28F \\ (10)
\textbf{31I},32Q,\textbf{33L},34D,38N,\textbf{39I},41Q,\textbf{42F},43N,\textbf{46E}
\\ (20)
49I,\textbf{50T},57V,61N,62F,\textbf{63F},\textbf{66V},\textbf{70T},72S,\textbf{75F}
\\ (30)
76Y,79F,83V,88L,90T,92F,\textbf{96F},\textbf{105V},106K,\textbf{107V}
\\ (40)
108H,109M,\textbf{110K},111K,\textbf{117S},\textbf{118Y},119W,\textbf{120V},\textbf{121F},122V
\\ \\ \textbf{Umbrella sampling.}  Umbrella
sampling~\cite{kn:chandler} was used to determine the number of states
$\Omega(R)$ for each of the 3 models.  First, different sterically
allowed conformations were collected for rms intervals of size 0.05
\AA\ over the entire rms range ($0-4.0$ \AA).  The rms values of
conformations differing only by random noise were averaged.  Next, the
probability to observe a conformation with rms $R$, $p_{i}(R)$, was
obtained for the $i$-th rms interval, $[R_i, R_{i+1}]$.  Since the
distribution of conformations $\Omega(R)$ must be a continuous
function of $R$, $\Omega(R)$ can be determined by appropriately
scaling the probabilities $p_i(R)$ to ensure continuity at interval
boundaries. Specifically, the probabilities $p_i(R)$ are sequentially
scaled by $p_{i-1}(R_i)/p_i(R_i)$, starting from $i = 2$.  $\Omega(R)$
is then obtained by multiplying all rescaled $p_i$'s by
$\Omega(R_1)/p_1(R_1)$.

The umbrella sampling scheme can only work provided that the
conformational space within each bin is explored evenly during
simulation.  In order to move the system into a particular rms bin, we
had to use $C=10$ and work at very low temperature ($T=0.005$).  These
conditions were necessary due to the large amount of entropy available
just outside any given bin, both in rms and in steric energy (the
number of clashed conformations is exponentially larger than the
number of well-packed ones).  The downside is that the system may end
up in a part of conformational space within a given bin that is
disconnected from other parts of the bin, i.e. the system behaves
non-ergodically under the conditions in which it must be sampled.  The
rms histograms within such isolated clusters are not representative of
the entire space within a bin, and can therefore lead to large errors
when the full rms histogram is assembled.  To overcome this
difficulty, we performed 200 short runs, starting from random
conformations, for each bin and computed the size of the cluster
explored in each run.  The cluster size is defined as the average
torsional distance between all pairs of uncorrelated conformations
encountered in a run.  We found that most bins had many clusters of
various sizes.  Since the number of states within a cluster should
scale exponentially with its size, the largest cluster encountered
should dominate the statistics within a given bin.  We therefore
sampled each bin by starting a run from a conformation obtained in a
short run that had entered the large cluster.  For every rms bin, we
collected 10,000 states in the largest cluster, and used only these
states when assembling the full rms histogram.

Several tricks proved handy when trying to find the large cluster.
The plot of the largest cluster's size over rms should be relatively
smooth.  If we found a cluster of size 20 at rms 0.90, a cluster of
size 10 at rms 0.95, and a cluster of size 24 at rms 1.00, we could be
certain that the 0.95 bin would be badly sampled.  We could thus focus
our efforts on bins with anomalously low cluster sizes.  Some bins
required up to 500 short runs before the large cluster was found.  We
observed that there is only one cluster at low rms values.  If we
started a run from the lowest rms bin, and allowed it to escape into a
higher rms bin, the run would always end up in the largest cluster for
that bin.  This worked up to rms 1.2 because entropy was steeply
increasing with rms, and one cluster was dominating the
statistics. For rms values between 1.2 and 2.0, several clusters of
similar sizes emerged, and thus sampling the largest one was no longer
adequate.  In this rms range, the distribution of states was
relatively flat over rms, and thus the entire range could be fully
sampled as a single large bin.  For rms higher than two, the method of
multiple short runs worked well.  The distribution of clusters as rms
is varied emerged as an interesting problem in itself, but was beyond
the scope of this paper.

\textbf{CHARMM minimization of structures} Using CHARMM, we minimized
and added hydrogen atoms to a set of 50 randomly chosen repacked
conformations of each of the three proteins used in this study.  All
such minimizations, including those performed for other randomly
selected packed structures, did not change the rotamer states of the
sidechains, but did eliminate close contacts (as defined by CHARMM)
resulting from the addition of hydrogens.  A set of 1000 repacked
conformations are available for public download at
http://www-shakh.harvard.edu/$\sim$shimada/index.html.

\section*{Results}

Figure \ref{fig:states} shows the distribution of \emph{packed}
(sterically allowed) conformations for three proteins, obtained over a
range of rms values using umbrella sampling.  For each protein, the
rms values are computed over the subset of partially occluded residues
(see Methods), and thus only the occluded positions contribute to the
total number of states.  All other positions are free to repack, but
are not counted in our sampling.  With \as $=$ \ah\ (the hard-sphere
model), the number of packed conformations of subtilisin is
approximately $10^{32}$.  As \as\ is increased using the soft-sphere
model, the number of packed conformations decreases.  This trend
reverses when the soft radii reach the van der Waals size (see
Discussion), at \as $= 1.0$, and the number of repackings is
approximately $10^{19}$.  We emphasize that the soft-sphere model
tolerates fewer clashes than the hard-sphere model, because it allows
no hard-sphere clashes (at \ah) and only a limited number of
soft-sphere clashes (at \as).  Using the steric LJ model, the number
of repackings is $10^{25}$.  Since the number of repackings scales
exponentially in the number of free torsions, we give the number of
rotamer states per torsion in Table 2 for each protein, calculated
from the LJ model.  These numbers are obtained by taking the $n$-th
root of the total number of states, where $n$ is either the number of
free torsions or residues.  Normalizing by the number of torsions
gives smaller variation over proteins than when normalizing by number
of residues, due to compositional differences between proteins
(variation in number of torsions per residue among proteins).

As a control, we applied the same sampling method to obtain the
distribution of states for all possible sidechain conformations,
packed and not packed, as shown in Figure \ref{fig:random_hists}.  The
total number of random sidechain conformations for a protein with $n$
free torsions is $3^n$.  For all three proteins, the area under the
random curve obtained from umbrella sampling is $n \log(3) \pm 2$,
demonstrating that the sampling is reliable, and that the total error
is approximately 2 orders of magnitude.  For each protein we included
the distribution of states using the hard-sphere model (dashed line)
in order to show the dramatic reduction in available conformations
once excluded volume constraints are applied.

In Figure \ref{fig:subtilisin_hists} we show the sampling of packed
conformations for subtilisin without umbrella constraints.  The same
data is shown separately over rms and over torsional distance from the
native conformation.  We see that the distributions over rms for the
different steric models often yield two humps.  When plotted over
torsional distance, all models give perfect normal distributions.  We
found no correlation between rms and torsional distance in the range
of rms that was sampled without umbrella constraints.  One
reason for this lack of correlation is that deviations in the
$\chi_{1}$ angle yield much larger deviations in rms than the other
$\chi$ angles.  Thus, unless rms is very low, it yields almost no
information about the torsional states of the protein.

The packing density of a conformation is usually defined as the volume
of the van der Waals envelope of a protein divided by its Voronoi
volume.  The density of a random sample of 1000 repacked conformations
of each protein was computed and found to vary by only $\pm 1$\% of
the native density.  Given this minimal variation in packing density,
we conclude that the requirement of having native-like density does
not significantly reduce the number of packed sidechain conformations.
In addition, we were able to build hydrogen atoms (using
CHARMM~\cite{kn:charmm}) on randomly selected, packed conformations,
confirming that the conformations were not too compact.

While the number of packed conformations is large, it is conceivable
that most repackings are structuraly similar to one another, and thus
that the number of truly different repackings may be much smaller than
observed.  To determine how different these repacked conformations
are, we uniquely identified each conformation by the rotamers of the
partially occluded positions.  The distance between two conformations
was measured by counting the number of single-bond torsions which had
to be rotated (by $120^{\circ}$) in order to obtain one conformation
from the other.  For example, for the partially occluded positions of
subtilisin, having a total of 174 torsions, we expect random
conformations to differ by 115 ($= 174 \times 2/3$) torsion moves.

The average torsional distance between repacked conformations (Fig.
\ref{fig:states}) shows that the conformations are different from each
other.  The similarity between conformations depends on the particular
model.  In general, larger soft radii lead to more similar repackings
until the soft radii reach the van der Waals size (see Discussion).
For the three proteins we examined, average conformations obtained
with the LJ model differ at 20\%-25\% of their torsions.  We note that
the differences between the soft-sphere models for \as $\geq 0.90$ and
the LJ model are minimal in all three proteins.
 
The entropy of each sidechain was computed by sampling uncorrelated
states meeting the LJ constraints for each protein
(Fig. \ref{fig:entropy}).  The plot shows that entropy is not
distributed evenly over positions.  Approximately 25\% of positions
have nearly zero entropy, implying that they are rarely repacked.  We
note that only 5\% of positions have exactly zero entropy,
demonstrating that almost all positions can accomodate more than one
rotamer.  The maximum entropy possible for a sidechain with $n$
torsions is $n\ln (3) = 1.1n$.  Since we observe positions with
entropy greater than 1.1, it is clear that large residues can take on
several different rotamers.

To explain the uneven distribution of entropy, we tried to correlate
it with various quantities.  The entropy of a given position showed no
correlation with either crystallographic B-factors, or with occluded
surface area.  We found significant but relatively low correlation
($R=0.65$) between the entropies we observed and entropies calculated
from rotamer probabilities from the PDB
database~\cite{kn:dunbrack_cohen}.  To investigate this trend further,
we plotted the rotamer probabilities observed in our simulation versus
probabilities observed in the pdb database.  Specifically, for each
rotamer state at every partially occluded position we plotted the
frequency with which it was observed in our simulation, versus the
frequency with which it occurs in the pdb database at its particular
$\phi$ and $\psi$ backbone angles.

These scatter plots for the three proteins are shown in Figure
\ref{fig:rotamer_probs}.  The plots indicate that native rotamers (red
marks, corresponding to the rotamers observed in the crystal
structure) are often seen with high probability in our simulations.
Since we observe in simulations only a few non-native rotamers (blue
marks) with probability greater than 90\%, the low entropy positions
in Figure \ref{fig:entropy} largely correspond to the native
conformation.  The pdb probabilities are obtained by averaging over
many different folds and sequences, and thus sequence-specific effects
are largely averaged out, leaving only the backbone $\phi$ and $\psi$
angles as a determinant of rotamer frequency.  Hence the simulation
frequency of rotamers lying near the diagonal is explained well by the
local backbone conformation at that position.  Rotamers lying off the
diagonal are being observed with anomalous frequency for their local
backbone conformation.  Because the backbone conformation does not
explain the frequency of these positions, they must be particularly
sensitive to one or more sequence-specific interactions.  We observe
several native rotamers with anomalously high probability (upper-left
corner), indicating that sequence-specific steric interactions can
completely stabilize the native rotamer even when it is rarely
observed in most other proteins.  We see a few non-native rotamers
anomalously frequently (blue marks in the upper-left of the figure).
These rotamers are preferentially selected by sequence-dependent
excluded volume interactions over the native rotamers.

One can classify positions, loosely speaking, as either
``well-predicted'' if the correct native rotamer is seen more than
50\% of the time, or ``poorly-predicted'' if the native rotamer is
seen less than 50\% of the time.  Non-native rotamers are plotted as
squares if they belong to ``well-predicted'' positions, and circles if
they belong to ``poorly-predicted'' positions.  A non-native rotamer
which is seen with high probability in the PDB database, may be seen
with low probability in our simulation, lying in the bottom-right
corner of the figures.  Most points in this region are squares, that
is, non-native rotamers which are being seen too infrequently because
they belong to ``well-predicted'' positions.  There are very few
circles in this region indicating that, for the most part, significant
deviations from observed pdb rotamer probabilities are due to
sequence-specific interactions stabilizing the native rotamer
preferentially over a more popular rotamer.

Concanavalin appears to have more scatter than the other two proteins.
In particular, it has more native rotamers below and near the
diagonal.  Concanavalin is an all-$\beta$\ protein, while subtilisin
is all-$\alpha$, and photoactive yellow protein is $\alpha$/$\beta$.
It is likely that the $\alpha$-helices of subtilisin, which act to
compactify the fold locally, impose more stringent constraints on the
allowed rotamers at a given position.  This is manifested in our
observation of fewer near-diagonal native rotamers in subtilisin than
in concanavalin.

To check whether simulation probabilities correlate with amino acid
type, we plotted the sequence identity of the native rotamers as
letters in Figure \ref{fig:native_types}.  We see that points lying in
the lower left-hand corner are generally polar or charged amino acids.
Native rotamers seen with high-probability in simulation are mostly
hydrophobic.  There are exceptions to both of these general trends.
Since hydrophobic residues are on average more buried than charged
residues, one might suggest that the separation based on polarity is
actually due to solvent-exposure.  As stated above, however, we found
no correlation between occluded surface and the simulation
probabilities.  In order to verify that this was not due to some
peculiarity of the definition of occluded surface, we visually
classified positions as buried or partially exposed using rough
criteria outlined in Methods.  The purple letters in Figure
\ref{fig:native_types} correspond to buried positions (as defined in
Methods), while the green letters correspond to partially exposed
positions.  We see no significant clustering of either color in these
figures, confirming that entropy in our simulations is not governed by
exposure.  We conclude, not surprisingly, that the importance of
electrostatic interactions, rather than solvent exposure, explains the
poorly-predicted native rotamers of charged amino acids.

We examined the correlation between the rotamer changes made at a pair
of positions $x$ and $y$ by measuring the number of states lost, $\exp
\left(\Delta S\right)$, due to interactions between $x$ and $y$, where
$\Delta S = S_{xy} - (S_{x}+S_{y})$, $S_{xy}$ is the joint entropy,
and $S_{x}$ and $S_{y}$ are the individual entropies.  We found only 3
pairs of positions in subtilisin exhibiting any correlation ($\exp
\left(\Delta S\right)$ between 0.1 and 0.3).  Similar numbers were
found in the other two proteins.  Repacking at a single position can
therefore be accomodated by the surrounding residues in many different
ways.

Given that there is a significant amount of torsional entropy in our
model, it interesting to ask how well an average sidechain structure
prediction based solely on excluded volume would perform.  For
subtilisin, we observe the rms of predictions broadly distributed
between 1.4 and 1.8 angsroms.  Using either the LJ model or the \as $=
0.90$ model, we predict correct rotamers for 65\% of residues, and
predict $\chi_{1}$ angles correctly for 85\% of positions.  The best
performing sidechain prediction algorithms, which incorporate several
physical and knowledge-based terms, predict 93\% of $\chi_{1}$ angles,
so excluded volume alone performs very well in this regard.

\section*{Discussion}
Upon folding to its native backbone topology, the conformational space
of a protein is cut by many orders of magnitude
(Fig. \ref{fig:random_hists}) due to excluded volume.  At the level of
a single torsion, the number of rotamer states is reduced by
approximately a factor two.  The native backbone can still accomodate
many different sidechain conformations, while maintaining its high
degree of compactness. Other sources of stabilization (such as
attractive van der Waals interactions, hydrogen bonding, polar
interactions) are necessary to overcome the final entropic cost
associated with fixing the sidechains in their native conformations.

The packing of sidechains in a native backbone has often been
described as a jigsaw-puzzle: the shapes of the sidechains (the
pieces) are enough to determine how they will fit in the protein
interior~\cite{kn:levitt_endgame}.  This argument has been used to
explain the high density observed in proteins.  The large number of
repackings and the diversity of these conformations
(Fig. \ref{fig:states}), all with native-like density, suggest that a
precise fitting of residues is not necessary for a high packing
density.  The rotamer frequency plots demonstrate that while there is
a sequence-specific excluded volume effect that acts to single out the
native rotamers, it is not sufficient to overcome the fact that the
backbone generally allows several different rotamers at each position.
The distribution of entropy shows that many positions admit
significantly more than one rotamer.  Furthermore, the correlation
calculations show that a conformational change at one position does
not necessitate a particular change at spatially proximate positions.
We conclude that while sidechains must fit together to avoid steric
clashes and maintain native-like density, these constraints alone do
not give rise to precise and unique fitting of residues.  Other
energetic interactions must bias sidechain conformations toward the
native state, leading to the observation of directional packing in
protein structures~\cite{kn:thornton_packing}.

The three models employed in our study yield consistent results.  All
three vastly reduce the total number of available states.  The
hard-shere model is consistently more forgiving than the other models.
The soft-sphere models with \as\ $\ge 0.90$\ all roughly agree to $\pm
2$ orders of magnitude.  The LJ model is only slightly more tolerant
than the soft-sphere models, suggesting that a two-shell model can
adequately model sterics without introducing continuous distance
dependence.  The soft-sphere model has two limiting behaviors.  For
\as = \ah, the soft-sphere model is identical to the hard-sphere
model.  For \as $\gg 1$, all atom pairs in the \emph{native} structure
are considered to clash, and thus the soft-sphere model will tolerate
all possible clashes.  Since the hard-sphere excluded volume is the
only remaining constraint, the soft-sphere model is again equivalent
to the hard-sphere model in this limit.  The observed rise in the
number of states as \as\ reaches 1.0 is the necessary turn-around
between these two limiting behaviors.  Sterics must therefore be
modelled with \as $< 1$.  When \as $= 1.0$, the radii have reached van
der Waals sizes, and are in the regime of attraction rather than
excluded volume.  While increasing the radii past the hard-core value
gives a significant gain in stability, we see that there is a limit on
how much stabilization can be achieved from excluded volume effects.

Our ability to predict 85\% of $\chi_{1}$ angles correctly shows that
excluded volume is of primary importance in determining the general
orientation of a sidechain with respect to its backbone.  Our
prediction of only 65\% of positions correctly shows that sterics are
less important in determining the rest of the sidechain's rotameric
conformation.  Of particular interest are native rotamers that are
observed in the PDB database in less than 10\% of proteins.  Many of
these rotamers are observed with significantly higher probability in
our simulation, while some are observed at their expected PDB level.
The excluded volume interaction is thus exhibiting some frustration
between the distribution of states compatible with the local backbone
conformation, and the sequence-specific neighborhood of the residue.
Other interactions must provide significant stabilization to overcome
this problem.

A natural extension of our work is to add an attractive interaction to
the potential which would stabilize the native packing.  From the
distribution of packed conformations, we estimate that the native
conformation must be about $0.6 kT$ per sidechain (see Table 2) lower
in energy than the average packed conformation with native backbone.
This estimate can be used to guide efforts towards designing better
potentials for sidechain packing.  Furthermore, the techniques
presented here, combined with a stepwise increase in the complexity of
potentials, can determine how much stabilization each additional
interaction provides.  It will be interesting to see whether a
potential that significantly stabilizes the native rotamers can
stabilize the native fold when backbone motions are added to the
model.

\newpage

\section*{Acknowledgements}

We are grateful for stimulating discussions with Gabriel Berriz,
Alexey Finkelstein, Leonid Mirny, and Michael Morrissey, and
especially thank Kirk Doran for his assistance.  Financial support
from NIH (grant 52126) and NSF (graduate fellowship, E.K.) is
acknowledged.

\newpage

\bibliographystyle{myjmb}
\bibliography{sidechain_packing}

\newpage
\section*{Figures}

\noindent
\underline{Figure 1} \\ The number of soft clashes in near-native
conformations as a function of \as\ for each protein studied.  The
solid line denotes the average number observed in an ensemble of
near-native conformations generated by adding $\pm 10^{\circ}$ random
noise to the native sidechains.  The dotted line is determined by
taking three standard deviations below the average, and corresponds to
the soft clash threshold used in this study.  \\

\noindent
\underline{Figure 2} \\ The figures in the left column show the
density of packed conformations as a function of rms for various
values of \as (indicated by the colored, solid curves).  The dashed
curve is the corresponding distribution using the repulsive
Lennard-Jones model.  Figures in the right column show the average
number of bond rotations needed to bring one conformation to another
as a function of rms for the various models.  The average is taken
over all pairs of 1000 randomly selected conformations found in a
given rms window.  \\

\noindent
\underline{Figure 3} \\ The distribution of random sidechain
conformations for each of the three proteins (solid lines).  Umbrella
sampling over rms without any excluded volume constraints was used to
construct these figures.  The dashed lines correspond to the
hard-sphere model (identical to the black curves from Figure
\ref{fig:states}) and are provided for comparison. \\

\noindent
\underline{Figure 4} \\ Uncorrelated packed conformations for
subtilisin sampled using the various models with no rms constraints
are histogrammed separately over rms from the native crystal
structure, and over torsional distance from the native rotamers.  Each
histogram is obtained over a total of 240,000 states.  The states
shown in this plot correpond to the tops of the rms histograms from
Figure \ref{fig:states}, but were obtained without umbrella sampling.
We therefore do not see any low rms states in these plots, because
there are exponentially fewer states as rms is lowered.  The native
rotamer configuration of subtilisin was obtained by minimizing rms in
simulation.  The rms of this nearly-native rotameric state is 0.5 \AA\
from the crystal structure.  The reason for the non-zero rms is that
deviations of more than 10 degrees from perfect rotamers are observed
in the crystal structure, and our simulations allows for no more than
10 degrees of tolerance around each rotameric value for a given
sidechain $\chi$ angle. \\

\noindent
\underline{Figure 5} \\ The entropy for the partially occluded positions
of each protein.  50,000 uncorrelated packed conformations were
sampled for each protein using the repulsive LJ model without umbrella
sampling.  The probability of observing each rotamer at each position
was calculated, and the entropy of each position was computed using $S
= -\sum p_i \log p_i$, where $p_i$ is the frequency of occurence of
the $i$-th rotamer at the given position.  The histograms of these
entropy values are also provided for each protein. \\

\noindent
\underline{Figure 6} \\ Observed simulation frequency of a rotamer
vs. the PDB database frequency of that rotamer, given its amino acid
identity and the $\phi$-$\psi$ angles of its backbone.  Database
probabilities were taken from the rotamer library described in
\cite{kn:dunbrack_cohen}.  Each point in the plots corresponds to a
single rotamer.  Red points are native rotamers (there is one red
point for each sequence position); blue points are non-native rotamers
(there are multiple blue points for each position).  We called
positions ``well-predicted'' if the native rotamer was observed more
than 50\% of the time, and ``poorly-predicted'' otherwise.  We further
classified the non-native rotamers (blue points) as squares if they
belonged to ``well-predicted'' positions or circles if they belong to
``poorly-predicted'' positions.  For example, if a native rotamer at a
given position is observed 85\% of the time, all the other, non-native
rotamers for that position will be squares.  Conversely, if at a
different position the native rotamer is seen only 15\% of the time,
all the non-native rotamers for that position will be circles.
Simulation frequencies are calculated over 50,000 packed conformations
obtained using the LJ model without umbrella sampling.

\noindent
\underline{Figure 7} \\ Simulation vs. PDB frequency for each
\emph{native} rotamer.  The native rotamers for each protein are
identified by their amino acid type (letter) and are colored purple if
they are in a buried position, and green if they are in a partially
exposed position.  The buried positions are given in the methods
section, indicated in bold type.
\newpage

\noindent
\section*{Tables} 
\begin{table}[h]
\begin{tabular}{|c|c|c|} \hline
atomic group & $r$ & \ah $r$ \\ \hline \hline
C3H0 & 1.61 & 1.21 \\
C3H1 & 1.76 & 1.32 \\ 
C4H1 & 1.88 & 1.41 \\ 
C4H2 & 1.88 & 1.41 \\ 
C4H3 & 1.88 & 1.41 \\ 
N3H0 & 1.64 & 1.23 \\ 
N3H1 & 1.64 & 1.23 \\ 
N3H2 & 1.64 & 1.23 \\ 
N4H3 & 1.64 & 1.23 \\ 
O1H0 & 1.42 & 1.07\\ 
O2H1 & 1.46 & 1.10 \\
S2H0 & 1.77 & 1.33 \\
S2H1 & 1.77 & 1.33 \\ \hline
\end{tabular}
\caption[]{}\label{table:atomic_radii} 
\end{table} 

\noindent
\underline{Table 1}\\
The atomic groups, VdW radii ($r$), and the hard core distances (\ah $r$) used in our study.  The atom groups and VdW radii were obtained from Table 2 of Tsai \emph{et al}. (1999).  The atomic group A$n$H$m$ refers to an atom of element A with a chemical valence of $n$ and $m$ hydrogen atoms bonded to it.  A methyl carbon ($-$CH$_{3}$), for example, falls under the C4H3 atomic group.

\newpage
\noindent
\section*{Tables} 
\begin{table}[h]
\begin{tabular}{|c|c|c|c|c|} \hline
protein & \# of $\chi$'s & \# of res. & States per $\chi$/per res. & S per $\chi$/per res.  \\ \hline \hline
1F91 & 99 & 49 (125) & 1.52 / 2.33 & 0.42 / 0.85 \\
1NLS & 151 & 88 (237) & 1.46 / 1.92 & 0.38 / 0.65 \\ 
1GCI & 174 & 100 (269) & 1.39 / 1.78 & 0.33 / 0.58 \\ \hline
\end{tabular}
\caption[]{}\label{table:protein_statistics} 
\end{table} 

\noindent
\underline{Table 2}\\ Statistics of repackings for photoactive yellow
protein (1F9I), concanavalin (1NLS), and subtilisin (1GCI) using the
repulsive Lennard-Jones model.  The second column lists the number of
torsions that contribute to the overall entropy.  These torsions
belong to residues having more than 40\% occluded surface area.  The
number of occluded residues having one or more free torsions is given
in the third column, with the total length of the protein in
parentheses.  The number of states and entropy (S) per $\chi$ and per
residue are calculated over the occluded residues only.

\pagestyle{empty}

\newpage
\begin{center}
\epsfig{file=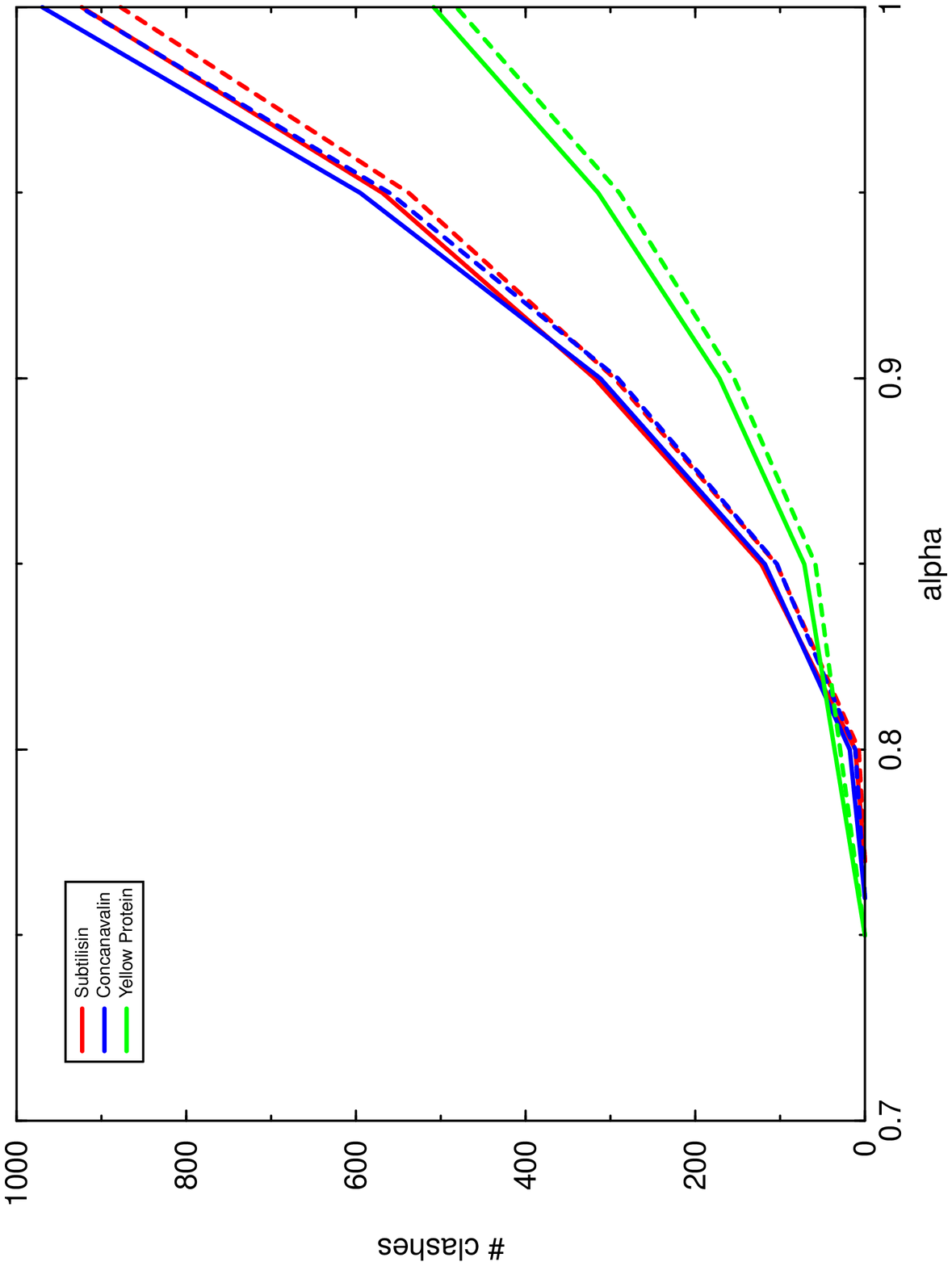}
\end{center}
\begin{center}
\epsfig{file=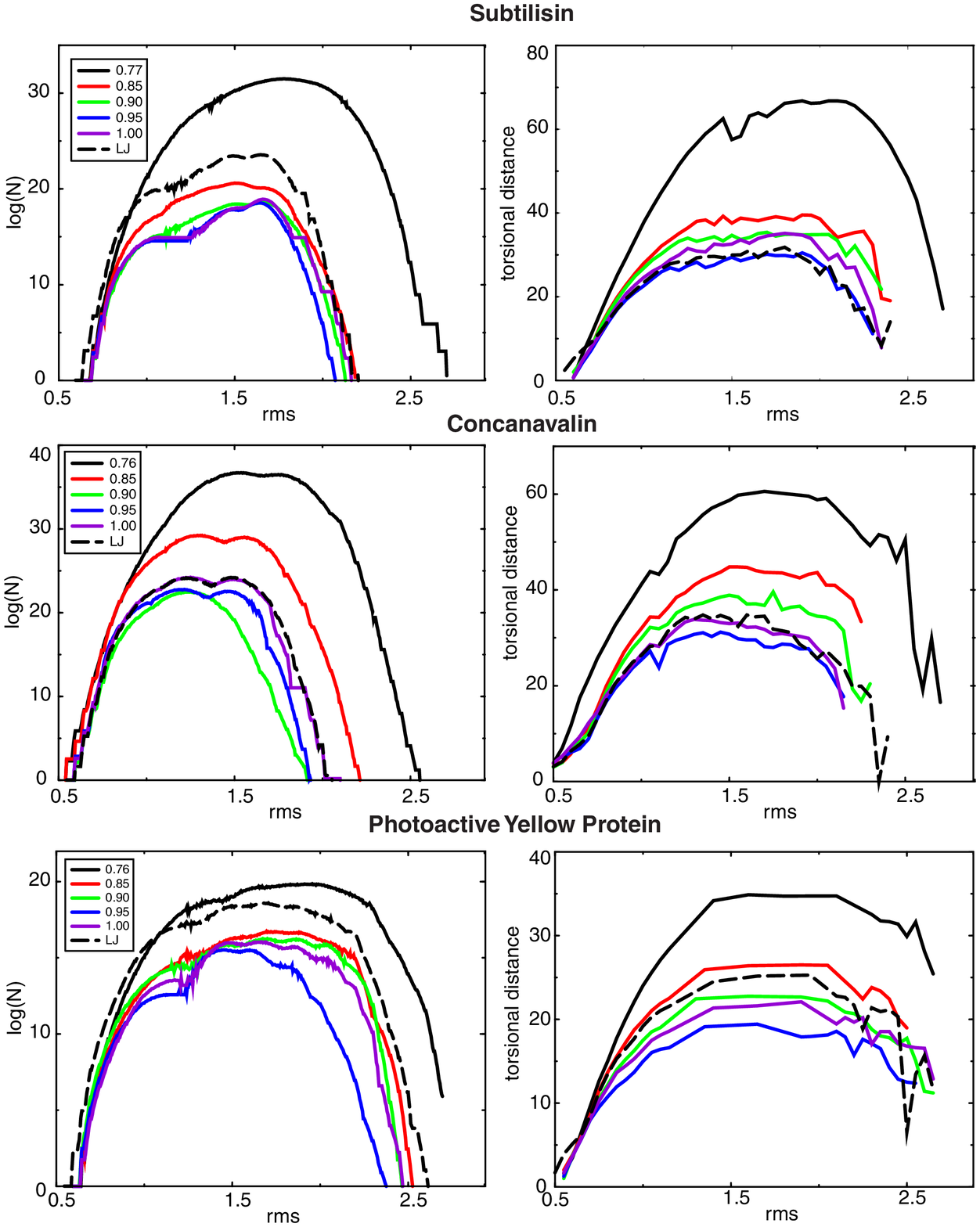}
\end{center}
\begin{center}
\epsfig{file=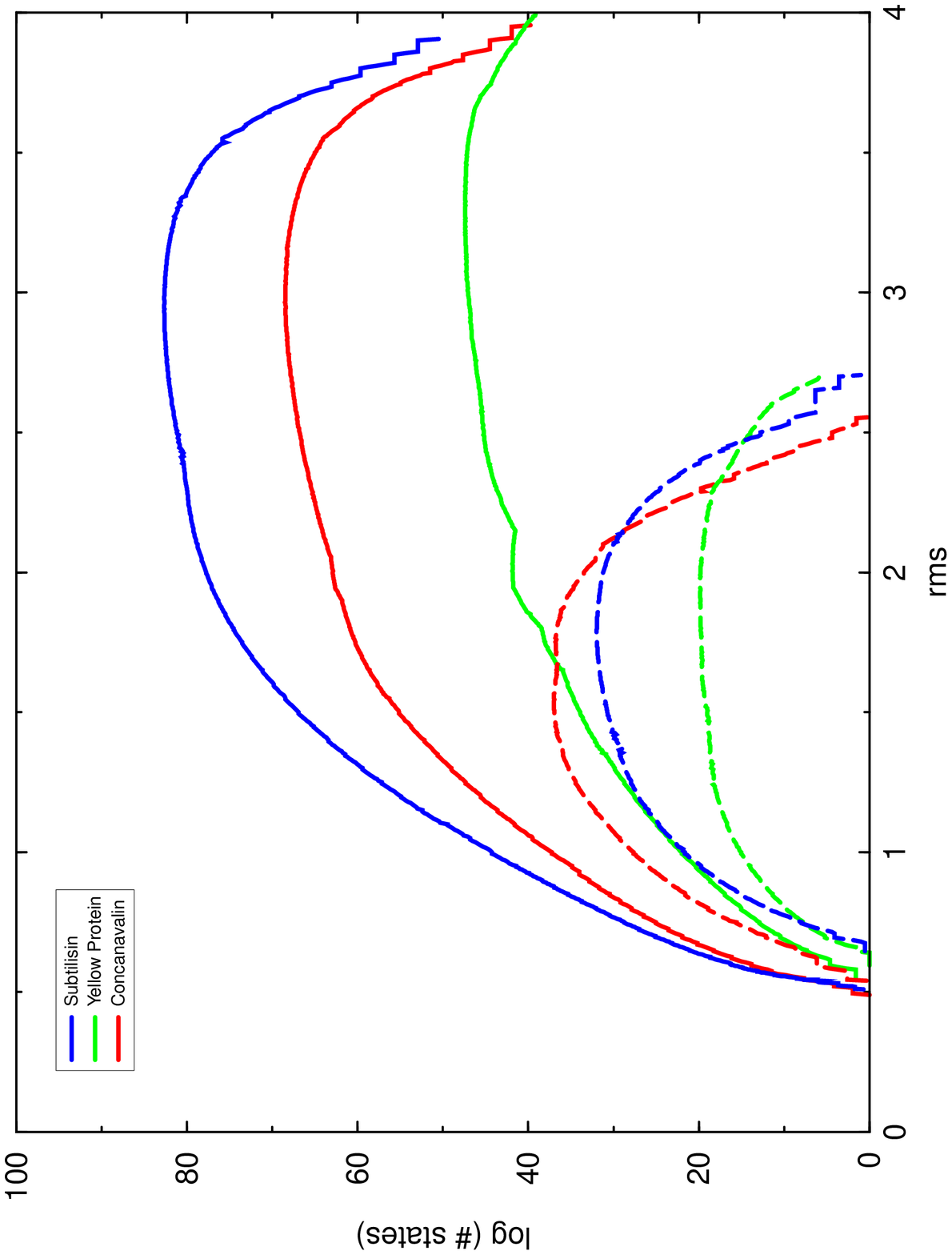}
\end{center}
\begin{center}
\epsfig{file=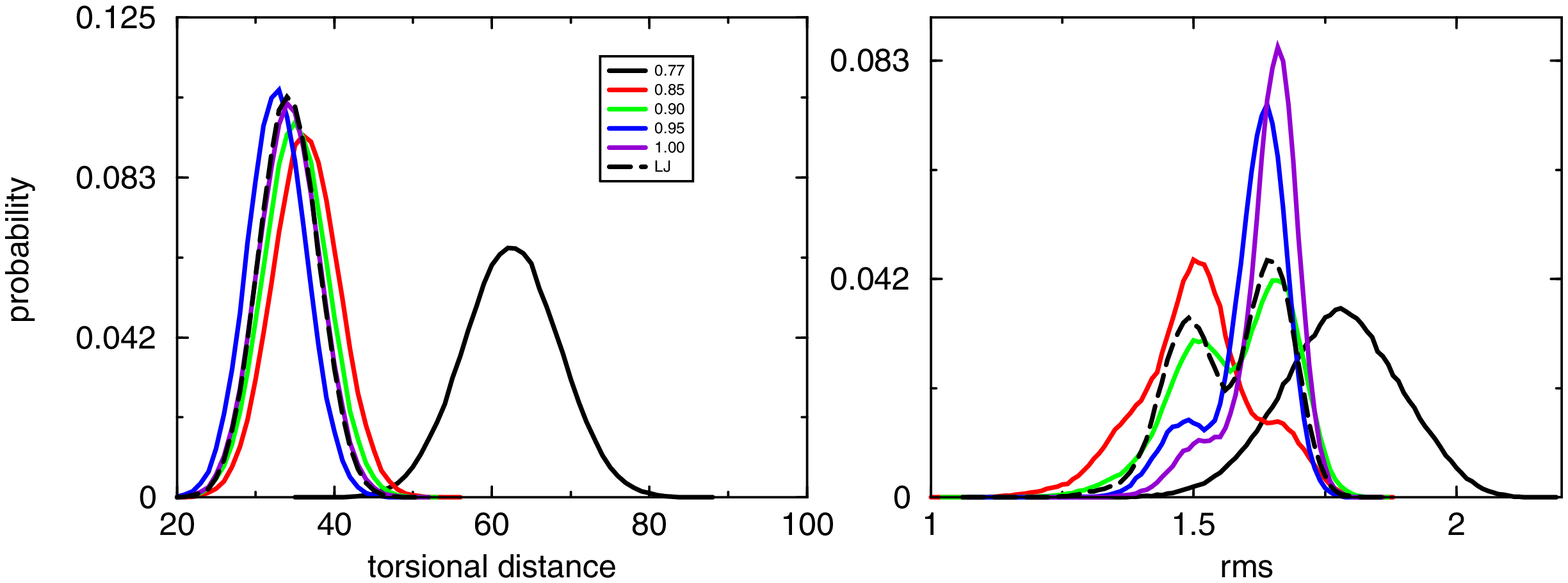}
\end{center}
\begin{center}
\epsfig{file=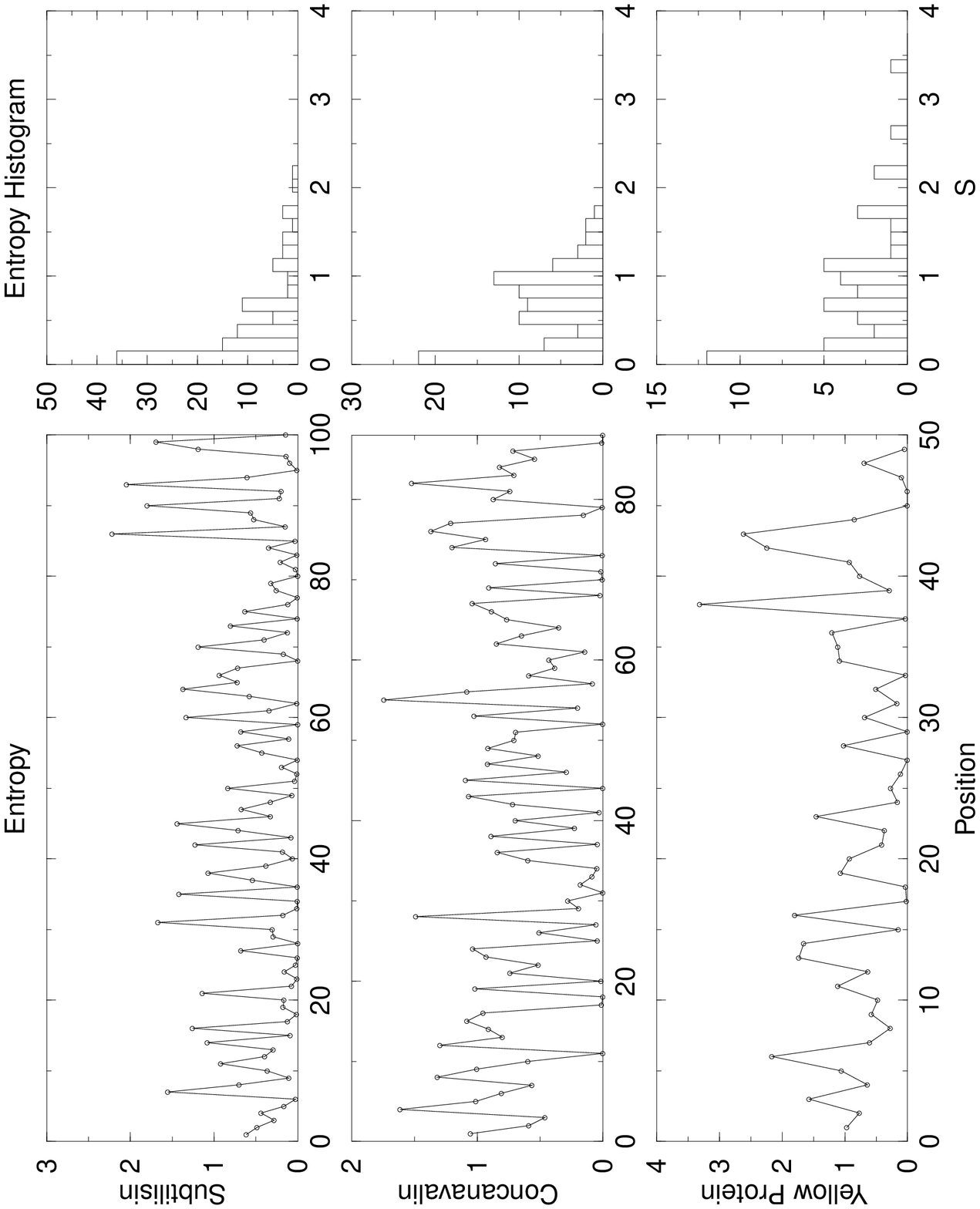}
\end{center}
\begin{center}
\epsfig{file=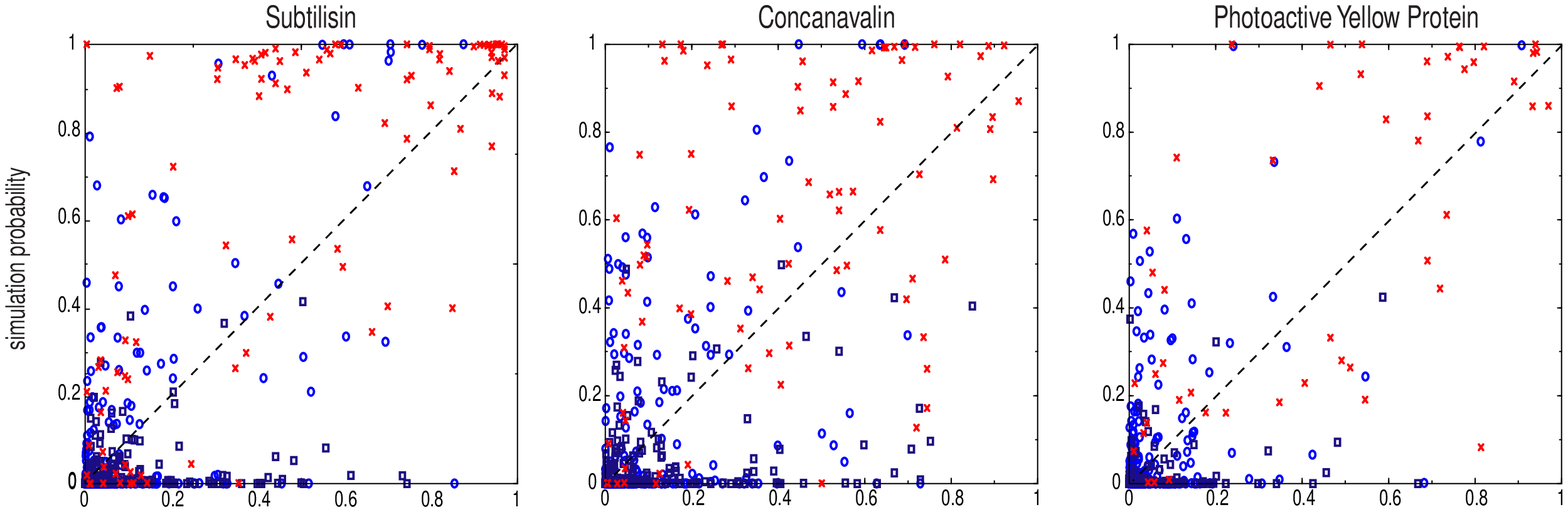}
\end{center}
\begin{center}
\epsfig{file=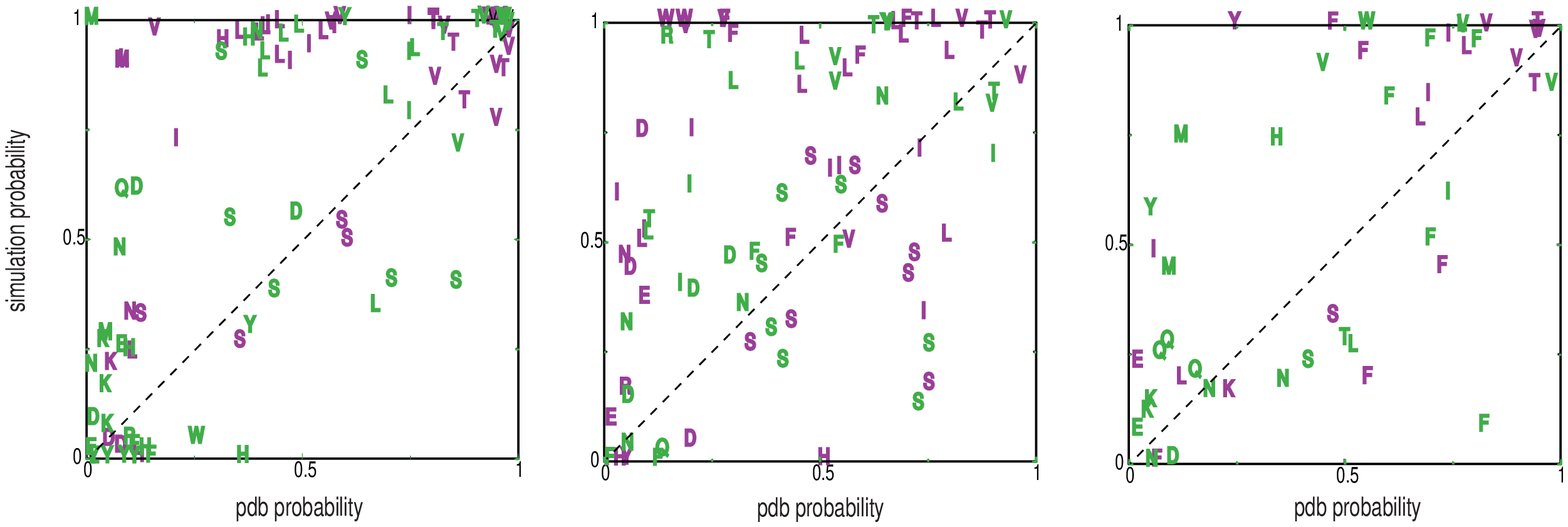}
\end{center}
\newpage

\begin{figure}[tbp]
\caption{} \label{fig:alpha_range}
\end{figure}
\begin{figure}[tbp]
\caption{} \label{fig:states}
\end{figure}
\begin{figure}[tbp]
\caption{} \label{fig:random_hists}
\end{figure}
\begin{figure}[tbp]
\caption{} \label{fig:subtilisin_hists}
\end{figure}
\begin{figure}[tbp]
\caption{} \label{fig:entropy}
\end{figure}
\begin{figure}[tbp]
\caption{} \label{fig:rotamer_probs}
\end{figure}
\begin{figure}[tbp]
\caption{} \label{fig:native_types}
\end{figure}
\end{document}